\documentclass[12pt]{article}
\usepackage{amsfonts}
\def\be{\begin{equation}}
\def\ee{\end{equation}}
\def\la{\label}
\def\ci{\cite}
\def\bi{\bibitem}
\def\<{\langle}
\def\>{\rangle}
\def\vf{\varphi}
\def\b{\begin{equation}}
\def\e{\end{equation}}
\def\be{\begin{eqnarray}}
\def\ee{\end{eqnarray}}
\def\R{\mathbb R}
\def\C{\mathbb C}

\begin{document}

\begin{center}

{\LARGE{\bf The Gamow Functional.}}   

\vskip1cm

{\large {\bf M. Castagnino$^\dagger$, M. Gadella$^{\dagger\dagger}$, R. Id
Bet\'an$^\dagger$, R. Laura$^\dagger$.  }} 

\end{center}

\vskip1cm
$^\dagger$Facultad de Ciencias Exactas, Ingenier\'{\i}a y Agrimensura. Av.
Pellegrini 250, (2000) Rosario, Argentina.  

$^{\dagger\dagger}$Departamento de F\'{\i}sica Te\'orica. Facultad de Ciencias.
c./ Real de Burgos, s.n. 47011 Valladolid, Spain.

\begin{abstract}

We present a formalism that represents pure states, mixtures and generalized
states as functionals on an algebra containing the observables of the system.
Along these states, there are other functionals that decay exponentially at all
times and therefore can be used to describe resonance phenomena.

\end{abstract}

\section{Introduction.}

Gamow vectors are vector states describing the exponentially decaying part of a
resonance \ci{G,B}.  Whether or not this exponentially decaying state is an
element of the reality is controversial \ci{FGR}. The question is if it can be
properly considered as a quantum state. 

If we assume that resonances are produced in a resonant scattering process
\ci{B,N}, produced by a ``free'' Hamiltonian $H_0$ and an interacting
Hamiltonian $H=H_0+V$,  Gamow vectors are eigenvalues of $H$ with complex
eigenvalues. This complex eigenvalues correspond to poles (that we are assuming
here to be simple) of the analytic continuation of the
$S$-matrix in the energy representation. These poles are located in the
lower half plane in the second sheet of the Riemann surface associated to the
transformation
$|p|=\sqrt E$. As the total Hamiltonian is self adjoint, the Gamow vector cannot
belong to a Hilbert space of states, but instead to the bigger space of a
Gelfand triplet or rigged Hilbert space (RHS), ${\bf\Phi}\subset{\cal
H}\subset{\bf\Phi}^\times$. As a consequence the Gamow vector, $|f_0\>$, cannot
be normalized. Worse of all, there is no clear manner of defining the mean value
of the energy on a Gamow vector \ci{CGI}.

Pure states as well as mixtures are represented by density  operators. In order
to study the properties of the Gamow ``state'', we may try to write it as a
density operator on an extended (or rigged) Liouville space. Here, new
difficulties arise that suggest that a Gamow vector cannot give rise a
reasonanble quantum state \ci{GL}.

Inspired by the methods of the Brussels school \ci{A1,PP}, the group of Rosario
has developed a formalism that allows us to make calculations using Gamow 
``states''
\ci{LC1,LC2,LC3,LC4,CL}. However, the use that has been made in these papers of
the Gamow object has been merely operational and no clear definition of it has
been provided. It is the aim of the present paper to give a possible definition
of the Gamow object as a functional on an algebra of states compatible with 
this formalism and give its most interesting properties.

\section{Algebras of observables and the Gamow functional.}

\subsection{Observables.}

Let us assume that $H_0$ has simple continuous spectrum equal to
$\R^+=[0,\infty)$. Then, for each $E\in\R^+$ there exists a generalized
eigenvector $|E\>$ of $H_0$: $H_0|E\>=E|E\>$. The vector $|E\>$ lies in the
bigger space of a RHS ${\bf\Phi}\subset{\cal H}\subset{\bf\Phi}^\times$. The set
of vectors $|E\>$ is complete in the sense that

\b
H_0=\int_0^\infty E\,|E\>\<E|\,dE \la{1}
\e
They also verify that $\<E|E'\>=\delta(E-E')$, where the Dirac delta is refered
to the integration from $0$ to $\infty$.

We say that an operator\footnote{Here, the operator $O$ maps $\bf\Phi$ into
${\bf\Phi}^\times$. This is a generalization of the usual concept of operator
that maps a subspace $\bf\Phi$ of the Hilbert space $\cal H$ into $\cal H$.}
$O$ is {\it compatible} with
$H_0$ if it can be written in the form:

\begin{equation}
O=\int_0^\infty dE\,O_E\,|E\rangle\langle E|+\int_0^\infty dE\int_0^\infty
dE^{\prime}\,O_{EE^{\prime}}\,|E\rangle\langle E^{\prime}|  \label{2}
\end{equation}
where $O_E$ is a function of the real variable $E$ and $O_{EE'}$ is a function
of the two dimensional variable $(E,E')$. 

We want that the set of operators compatible with $H_0$ form an involutive
algebra with identity. In order to do this, we need to make a choice on the
functions $O_E$ and $O_{EE'}$.  Let $\cal D$ be the space of infinitely
differentiable functions, at all points, with compact support. This space is
endowed with a locally convex topology \ci{R}. Now, take the Fourier transform of
all functions in $\cal D$. The resulting space, ${\cal Z}\equiv {\cal F}({\cal
D})$, is a vector space of entire analytic functions with these two properties:

i.) The product of two functions $f(z),g(z)\in\cal Z$, $f(z)g(z)$, is also in
$\cal Z$.

ii.) If $p(z)$ is a polynomial and $f(z)\in\cal Z$, then, $p(z)f(z)$ is also in
$\cal Z$.

Then, we choose $O_E$ as the sum of a polynomial on $E$ plus a
function\footnote{Or the restriction to $\R^+$ of a function in $\cal Z$. Due to
the fundamental theorem of analytic continuation (see \ci{MA}), this restriction
fix uniquely the function in $\cal Z$.} in
$\cal Z$ (one or the other or both may be eventually equal to zero). The two
variable function $O_{EE'}$ belongs to the algebraic tensor product ${\cal
Z}\otimes {\cal Z}$, i.e., it must be of the form

\b
O_{EE'}=\sum_{ij} \lambda_{ij}\,\psi_i(E)\,\phi_j(E') \la{3}
\e
where $\psi_i(E),\,\phi_j(E')\in \cal Z$. The sum in (\ref{3}) is finite. The set
of operators compatible with $H_0$ forms an algebra where the sum of two
operators and the product by scalars are given in an obvious manner. To multiply
two operators, we must take into account that  $\<E|E'\>=\delta(E-E')$ and i.)
and ii.) as above. Then, this set form an algebra we call ${\cal A}_0$. This
algebra can be endowed with a topology due to the fact that the space $\cal P$
of polynomials can be endowed with a topology and  $\cal Z$ and ${\cal
Z}\otimes {\cal Z}$ have natural topologies \ci{R,P}. Thus, ${\cal A}_0$ is
isomorphic (as a vector space) to the direct sum ${\cal P}+{\cal Z}+{\cal
Z}\otimes {\cal Z}$.  

Next, let us assume that the choice of $(H_0,H)$ (or $V$) is such that the
M{\o}ller wave operators exist and the scattering is asymptotically complete.
Then, we can define the vectors\footnote{These vectors belong to the bigger space
in the RHS ${\bf\Phi}^\pm\subset{\cal H}\subset ({\bf\Phi}^\pm)^\times$, where
${\bf\Phi}^\pm={\bf\Omega}_\pm\bf\Phi$ \ci{BG}.}
$|E^\pm\>:={\bf\Omega}_\pm|E\>$ that are generalized eigenvectors of the total
Hamiltonian: $H|E^\pm\>=E|E^\pm\>$. If $O$ is an observable compatible with
$H_0$, let us write:

\be
O^\pm:&=& {\bf\Omega}_\pm\,O\,{\bf\Omega}_\pm^\dagger \nonumber\\[2ex] &=&
\int_0^\infty O_E\,|E^\pm\>\<E^\pm|\,dE + \int_0^\infty dE \int_0^\infty dE' \,
O_{EE'}\,|E^\pm\>\<E'^\pm| \la{4}
\ee

It is natural to say that an operator is compatible with $H$ if and only if it
has one of the forms given in (\ref{4}). Since

\b
\<E^\pm|w^\pm\>=\<E|\mathbf{\Omega }_{\pm }^{\dagger }\, \mathbf{\Omega }_{\pm
}|w\>=\<E|w\>=\delta(E-w) \la{5}
\e
the set of operators of the form (\ref{4}) form two algebras, ${\cal A}_\pm$,
isomorphic to
${\cal A}_0$. The topology on ${\cal A}_\pm$ is defined exactly as for ${\cal
A}_0$.

These algebras are involutive. If $O\in {\cal A}_0$, its adjoint is

\begin{equation}
O^{\dagger }:=\int_{0}^{\infty }dE\,O_{E}^{*}\,|E\rangle \langle
E|+\int_{0}^{\infty }dE\int_{0}^{\infty }dE^{\prime }\,O_{EE^{\prime
}}^{*}\,|E^{\prime }\>\langle E|  \label{6} 
\end{equation}
It is easy to show that this definition is consistent with the formula $
(\varphi ,O\psi )=(O^{\dagger }\varphi ,\psi )$, when $\varphi ,\,\psi \in 
\mathbf{\Phi }$,  where $\mathbf{\Phi }$ is a suitable space of test vector on
which $|E\>$ acts \ci{BG}. Similar definition applies for the adjoint of
$O^\pm$. An operator for which $O=O^\dagger$ is called  self adjoint. $O$ (or
$O^\pm$) is self adjoint if and only if
$O_E=O^*_E$ and $O_{EE^{\prime}}=O^*_{E^{\prime}E}$. Self adjoint members of
${\cal A}_0$ (${\cal A}_\pm$) are the observables of the system. If we write 

\b
|E^\pm):=|E^\pm\>\<E^\pm| \hskip0.5cm;\hskip0.5cm |EE'^\pm):= |E^\pm\>\<E'^\pm|
\la{7}
\e
then, $O^\pm$ can be written as:

\b
O^\pm= \int_0^\infty dE\,O_E\,|E^\pm)+ \int_0^\infty dE \int_0^\infty dE'
\,O_{EE'} \,|EE'^\pm) \la{8}
\e

\subsection{States.}

Let us consider the duals, ${\cal A}_\pm^\times$, of the algebras ${\cal
A}_\pm$. These are the vector spaces of continuous linear functionals
(functionals are mappings from ${\cal A}_\pm$ into the set $\C$ of complex
numbers) on
${\cal A}_\pm$. Examples of vectors on ${\cal A}_\pm^\times$ are the mappings

\b
O^\pm\longmapsto O_E\hskip0.5cm;\hskip0.5cm O^\pm\longmapsto O_{EE'} \la{9}
\e
for given $E$ and $E'$ in $\C$. When $E$ and $E'$ are real these mappings are
called $(E^\pm|$ and $(EE'^\pm|$ respectively, so that

\b
(E^\pm|O^\pm)=O_E \hskip0.5cm;\hskip0.5cm (EE'^\pm|O^\pm)=O_{EE'} \la{10}
\e

If $\rho^\pm$ is a linear functional on ${\cal A}_\pm$, we denote the action of
$\rho^\pm$ on $O^\pm$ by $(\rho^\pm|O^\pm)$.

It is customary to define a state on ${\cal A}_\pm$ as a continuous linear
functional
$\rho^\pm$ on ${\cal A}_\pm$ verifying the following conditions \ci{BR}:

\smallskip
i.) Positivity: $(\rho^\pm|(O^\pm)^\dagger\, O^\pm)\ge 0$ for all $O^\pm\in {\cal
A}_\pm$.

ii.) Normalization: $(\rho^\pm|I^\pm)=1$, where $I^\pm$ are the identities on
the algebras ${\cal A}_\pm$. These identities can be written as\footnote{The
operators $I^\pm$ are the canonical imbeddings of ${\bf\Phi}^\pm$ into their
respective dual spaces $({\bf\Phi}^\pm)^\times$, i.e., $I^\pm\vf^\pm=\vf^\pm$,
see
\ci{GA}.}

\b
I^\pm=\int_0^\infty dE\,|E^\pm\>\<E^\pm|= \int_0^\infty dE\, |E^\pm). \la{11}
\e

In general a functional  on ${\cal A}_\pm$ can be written as 

\begin{equation}
\rho^\pm =\int_{0}^{\infty }dE\,\rho _{E}\,(E^\pm|+\int_{0}^{\infty
}dE\int_{0}^{\infty }dE^{\prime }\,\rho _{EE^{\prime }}\,(EE'^\pm|
\label{12}
\end{equation}
where $\rho_E$ and $\rho_{EE'}$ can be either functions or distributions, so that
when applied to
$O^\pm$ the result is

\begin{equation}
(\rho^\pm |O^\pm)=\int_{0}^{\infty }dE\,\rho
_{E}\,O_{E}+\int_{0}^{\infty }dE\int_{0}^{\infty }dE^{\prime }\,\rho
_{EE^{\prime }}\,O_{EE^{\prime }}  \label{13}
\end{equation}
observe that (\ref{13}) implies the relations

\b
(E^\pm|w^\pm)=\delta(E-w) \hskip0.5cm;\hskip0.5cm
(EE'^\pm|ww'^\pm)=\delta(E-w)\,\delta(E'-w').
\la{14}
\e

The evolution of the states $\rho^\pm$ under the total Hamiltonian $H$ can be
defined using the following expression:

\be
(\rho^\pm_t|O^\pm)&=&(\rho|e^{itH}\,O^\pm\,e^{-itH}) \nonumber\\[2ex]
&=&\int_0^\infty dE\, \rho_E\,O_E + \int_0^\infty dE\,dE'
e^{it(E-E')}\,\rho_{EE'}\,O_{EE'} \la{15}
\ee
which gives the evolution of the mean values using the Heisenberg definition of
the evolution for the observables. The Schr\"odinger evolution of the states then
results from (\ref{15}) and is
\b
\rho^\pm_t= \int_0^\infty dE\, \rho_E\,(E^\pm| + \int_0^\infty dE\,
\int_0^\infty dE'\, e^{it(E-E')}\,\rho_{EE'}\,(EE'^\pm| \la{16}
\e
We see that the first term in the right hand side of (\ref{16}) (the diagonal
part of $\rho_t^\pm$) is invariant and the second term (the nondiagonal part)
evolves with time. In accordance with this result, diagonal and nondiagonal parts
of (\ref{16}) are often called the invariant and the fluctuating parts
respectively \ci{LC1,CL}.

Normalized vectors on Hilbert space representing quantum pure  states and
positive trace one operators representing mixtures can be easily written in the
form (\ref{12}). If $\psi(E)$ represents the wave function of a pure state in the
energy representation, the coefficients $\rho_E$ and $\rho_{EE'}$ in (\ref{12})
are given by

\b
\rho_E=|\psi(E)|^2 \hskip0.5cm{\rm and}\hskip0.5cm
\rho_{EE'}=\psi^*(E)\,\psi(E') \la{17}
\e

A mixture can be written in the form $\rho =\sum_{i}\lambda_{i}\,|\psi
_{i}\>\langle \psi _{i}|$ with $\<\psi_i|\psi_j\>=\delta_{ij}$ and
$\sum_i\lambda_i=1$. If $\psi_i(E)$ is the wave function of the state $\psi_i$
in the energy representation, we have in here:

\b
\rho_E= \sum_{i}\lambda _{i}\,|\psi_{i}(E)|^{2} \hskip0.5cm{\rm and}\hskip0.5cm
\rho_{EE'}= \sum_{i}\lambda _{i}\,\psi _{i}^{*}(E)\,\psi _{i}(E^{\prime}) \la{18}
\e
Observe that, in any case, $\rho_E=\rho_{EE}$. In addition to pure states and
mixtures, this formalism allows for another kind of states called the states
with diagonal singular or generalized states \ci{A1} for which
$\rho_E\ne\rho_{EE}$.

\subsection{Gamow functionals.}

Gamow vectors are usually obtained as residues of poles in analytic
continuations of the $S$ operator (in the energy or momentum representations)
\ci{B,BG} or a reduced resolvent \ci{AP,AM,E}. In previous works \ci{LC1,CL},
Gamow vectors have been heuristically obtained  using analytic continuations on
the variables
$E$ and $E'$ of the fluctuating part of $(\rho_t^\pm|O)$, i.e., the second term
in the right hand side of (\ref{15}) and then, using a limiting process. Only
the fluctuating part is there used because this is the only relevant to time
evolution. Then, the result obtained does not have diagonal part. This suggest
that a good definition of the Gamow vector should have diagonal part equal to
zero.

Now we are in the position of defining the decaying Gamow functional in the
following form:

\b
\rho_D:= \int_0^\infty dE\int_0^\infty dE'\,\delta_{z_0^*}\otimes
\delta_{z_0}\, (EE'^+| \la{19} 
\e 
Observe that, in our definition, the component $\rho_E$ of $\rho_D$ vanishes and
the component $\rho_{EE'}$ is equal to $\delta_{z_0^*}\otimes
\delta_{z_0}$. When applied this distribution to a given $O_{EE'}\in{\cal
Z}\otimes {\cal Z}$, we obtain $O_{z_0^* z_0}$, i.e., the value of the function
$O_{EE'}$ at the point $(z_0^*,z_0)$. Obviously 

\b
(\rho_D|O^+)= O_{z_0^* z_0} \la{20}
\e

The decaying Gamow functional has the following properties:

1.- It is a continuous linear functional on ${\cal A}_+$.

2.- It is positive: $(\rho_D|(O^+)^\dagger\, O^+)\ge 0$.

3.- It decays exponentially for all values of time. If we write
$\rho_D(0)=\rho_D$, then,

\begin{eqnarray}
&&(\rho _{D}(t)|O^+): =(\rho_D|e^{itH}\,O^+\,e^{-itH}) \nonumber \\[2ex]
&=&\int_{0}^{\infty }dE\int_{0}^{\infty }dE^{\prime }\,\delta
_{z_{0}^{*}}\otimes\delta_{z_{0}}\,\,O_{EE^{\prime }}\,e^{it(E-E^{\prime
})}=e^{it(z_{0}^{*}-z_{0})}\,O_{z_{0}^{*}\,z_{0}}  \nonumber \\[2ex]
&=&e^{-t\Gamma }\,O_{z_{0}^{*}\,z_{0}}=e^{-t\Gamma }\,(\rho _{D}|O^+) \la{21}
\end{eqnarray}
It is not difficult to show, using the inverse Fourier transform, that
$O_{EE'}\,e^{it(E-E')}$ is in ${\cal Z}\otimes {\cal Z}$ for all values of $t$.
In fact, if $\vf(E)\in\cal Z$, we have that

\b
{\cal F}^{-1} (e^{itE}\,\vf(E))=\int_{-\infty}^\infty e^{itE}\,e^{i\tau
E}\,\vf(E)\,dE = ({\cal F}^{-1}\vf)(t+\tau) \la{22}
\e
As $({\cal F}^{-1}\vf)(\tau)$ is an infinitely differentiable function with
compact support, so is (\ref{22}).

The exponential decay of $\rho_D(t)$ for all values of time allows us the call
it the decaying Gamow functional. Observe that its decay mode is not a
semigroup and this avoids the difficulties in determining the instant $t=0$ at
which ``the preparation of the  quasistationary state is completed and starts
to decay'' \ci{B,BAK}. Nevertheless, a redefinition for $O_E$ and $O_{EE'}$ is
possible so that $\rho_D(t)$ is defined and decays exponentially for $t\ge 0$
only \ci{CGIL}. 

4.- Cannot be normalized: $(\rho_D|I^+)=0$. This means that $\rho_D$ does not
fit with the traditional definition of state \ci{BR}. Although most of data
on unstable particles seems to confirm the exponential decay \ci{PDG}, 
traditional quantum mechanics forsees deviations of the exponential decay law
for very small and very large times \ci{FGR}. Traditionally, Gamow vectors and
functionals have been used to construct generalized spectral expansions \ci{GA},
useful for calculations \ci{LC1,LC2,LC3,LC4,CL}.

5.- It is also obvious that

\begin{equation}
(\rho _{D}|H^n)=0, \hskip0.7cm n=0,1,2,\dots \la{23}
\end{equation}
where

$$
H^n=\int_0^\infty dE\,E^n \,|E^+)
$$
For $n=0$, we recover $(\rho_D|I^+)=0$. For $n=1$, we obtain $(\rho _{D}|H)=0$.
This means that the mean value of the energy on the Gamow state is zero, which
agrees with some predictions \ci{PPT} but contradicts some others \ci{CGI}. 
In our opinion, (\ref{23}) is an argument against the option of considering
$\rho_D$ as a physical state, as all momenta of $H$ on $\rho_D$
vanish.

\vskip1cm
{\bf \Large Concluding remarks}

\medskip
We have presented here a formalism in which pure states, mixtures and generalized
states with singular diagonal \ci{A1} are positive, normalized continuous
functionals on an algebra containing the observables of a quantum system. Along
these functionals we have obtained a functional that decays exponentially, the
Gamow functional, but for which no normalization exists. This functional is
always present in decaying processes and is useful for  calculations.
However, the  fact that the mean value of 
$H^n$ for
$n=0,1,2,\dots$ in
$\rho_D$ is always zero, suggest that the Gamow functional cannot be taken
seriously as a truly physical state.

\section*{Acknowledgements} We thank Drs. I.E. Antoniou, A. Bohm, R. de la
Madrid, L. Lara and A. Ord\'o\~nez for enlightening discussions. Partial
finantial support is acknowledged to DGICYT PB98-0370, DGICYT PB98-0360, the
Junta de Castilla y Le\'on Project PC02/99, the CONICET and the National
University of Rosario.

\end{document}